\documentclass{article}   	

\usepackage{graphicx}						
\usepackage{amssymb}
\usepackage{amsmath}
\usepackage{mleftright}

\usepackage{physics}
\usepackage{autonum}
\usepackage{float}

\usepackage{cite}

\usepackage[T2A]{fontenc}
\usepackage[utf8]{inputenc}
\usepackage{braket}

\usepackage[titletoc,title]{appendix}

\begin{document}

\begin{titlepage}
 	  \begin{center}

    		   \huge{Eﬀect of initial intrasystem
entanglement on entropy growth in
generalized Jaynes–Cummings models}

			\vspace{1cm}

		\Large{Daria Gaidukevich}
			\vspace{0.3cm}

		\large{Independent researcher}
		\vspace{0.3cm}
		
		\large{dasha.gaidukevich@gmail.com}
		\vspace{0.3cm}

 	  \end{center}

\begin{abstract}
We investigate how initial intrasystem entanglement influences the entropy generated in atomic systems interacting with a photonic environment in several generalizations of the Jaynes–Cummings model with two or more subsystems. Since the initial entanglement does not uniquely determine the final entropy, we focus on ensemble-averaged behavior. We consider ensembles of initial system states, including pure and mixed Haar-random states, ensembles with fixed average energy or fixed mixedness, and varying initial photon numbers in the environment. In all cases, we observe a positive correlation between the initial entanglement and the entropy growth, although the fractional contribution of the initial entanglement varies. Our results emphasize the role of intrasystem correlations as a factor contributing to entropy growth in quantum informational processes.

\end{abstract}

\end{titlepage}

 \section{Introduction}

 The interaction of quantum systems with their environment typically leads to information transfer, a subject of interest not only in quantum information theory but also in related fields such as black hole physics and quantum thermodynamics. The flow of different types of correlations into one another as well as restrictions that they impose on each other have been discussed in the literature (see, e.g., Refs.~\cite{Distributed, Monogamy, transfer, Overview, Quantitative, Redundantly, Darwinism}). This naturally raises the question of whether stronger initial internal correlations lead to the generation of stronger system–environment correlations during the interaction.
 
  Indirect evidence for a positive answer appears in Refs.~\cite{Dynamics, Flow, Distributed, Protecting, Decoherence}. In particular, by examining the plots of entropy dynamics in Ref.~\cite{Dynamics}, one can see that, on average, maximally entangled initial states evolve into states with higher von Neumann entropy than separable ones.

  The issue of the influence of initial intrasystem entanglement on entropy generation was raised in Ref.~\cite{Does} and examined using the example of a system of qubits interacting with an environment described by the quantum harmonic oscillator. It was shown that although the relationship between the two quantities is not straightforward, the average dependence for Haar-random states is positive. This behavior was attributed to the concentration-of-measure phenomenon. At the same time, the dependence was found to vary strongly between ensembles, and for small systems one can find examples where the average dependence becomes negative. That study, however, considered only pure initial states and a single model, from which the role of the environment’s initial state could not be inferred.

Here, we extend this investigation using generalizations of the Jaynes–Cummings model. The original model describes the interaction of a two-level atom with a single mode of the electromagnetic field  \cite{Comparison}. As a starting point, we consider the double Jaynes–Cummings setup consisting of two noninteracting subsystems with initially entangled atoms. This model has been considered in Refs. \cite{Entanglement, Atom-field, Sudden, Masood}. Ref.~\cite{Atom-field} investigated system-environment entanglement dynamics for both entangled and unentangled initial states, but the effect of initial entanglement on entropy was not addressed.

In this work, we focus on atomic systems and treat the fields as an environment. We show that, for each model generalization and each initial-state ensemble considered, initial intrasystem entanglement, on average, enhances the time-averaged entropy change of the system. We further analyze how this dependence varies across different samples of initial system states, including not only pure, but also mixed ensembles. In addition, we consider the effect of the environment’s initial photon number. We identify parameters that enhance or suppress the role of initial entanglement in entropy generation.  

\section {Model and Setup}

 The Hamiltonian of the standard Jaynes–Cummings model has the form (here and throughout, we set $\hbar = 1$):
\begin{equation}
H=H_0+H_{AF}=\frac{\omega_A}{2}r_{3}+\omega(a^\dagger a+\frac{1}{2})+g(ar^++a^\dagger r^-),
\label{Ham}
\end{equation}
where $g$ is the coupling constant defining the atom–field interaction, $\omega_A$ is the atomic transition frequency, $\omega$ is the field mode frequency,  $a^\dagger, a$ are the field-mode creation and annihilation operators, $r^+=\ket{e}\bra{g},r^-=\ket{g}\bra{e}$ are the atomic raising and lowering operators, and $r_3=\ket{e}\bra{e}-\ket{g}\bra{g}$, with $\ket{e}$ and $\ket{g}$ denoting the excited and ground atomic states, respectively.  Further details, including the explicit form of the eigenvectors and eigenvalues, can be found in Ref.~\cite{coherent}. In what follows, we describe the states of the atom–field system using the notation $\ket{g,m}$ and $\ket{e,m}$, where the second symbol denotes the photon number.

We consider generalizations of the model consisting of $N$ non-interacting Jaynes–Cummings subsystems, which we denote by $N\text{-JC}$.  Eigenvectors and eigenvalues of the corresponding Hamiltonian are given by all possible combinations of the eigenvectors $\ket{v_j}$ and eigenvalues $a_j$ of individual subsystems: $\ket{v_{j_1 \dots j_N}} =\ket{v_{j_1}}\otimes \ldots \otimes \ket{v_{j_N}}$ and $ a_{j_1\dots j_N}=a_{j_1}+ \ldots + a_{j_N}$, respectively. Initial states $\ket{\psi(0)}$ are expanded in the eigenbasis and evolve as: 
 \begin{equation}
 \sum_{j_1 \dots j_N}\braket{v_{j_1 \dots j_N}|\psi(0)}\ket{v_{j_1 \dots j_N} }e^{-it a_{j_1 \dots j_N}}. 
 \label{evolnq}
  \end{equation}
  To describe the evolution of a mixed initial state, one may generalize expression (\ref{evolnq}); alternatively, in the both cases one can switch to the eigenbasis of the Hamiltonian and define the components of the time-evolved density matrix as 
  \begin{equation}
\rho_{ij}(t)=e^{-i t (E_i-E_j)}\rho_{ij}(0), 
 \label{evolmix}
  \end{equation}
 where $E_i$ are the corresponding eigenvalues of the Hamiltonian. 

Tracing out the field degrees of freedom yields the reduced density matrix of the atomic system, for which the von Neumann entropy is computed. Throughout our analysis, we employ the time-averaged von Neumann entropy, with logarithms taken in base two, denoted by $S_t$. In what follows, the overline denotes averaging over the ensemble.

In Secs.~3, 4, and 6, we consider a model with two-level atoms, where each subsystem is described by the Hamiltonian (1). In this case, the system dynamics depends on the atom–field detuning $\delta=|\omega_A-\omega|$ and the coupling $g$. Since our analysis involves time averaging, the results depend only on the ratio of these parameters. In all calculations, we set $\delta=2g$. This choice is motivated by the observation that, both in the dispersive regime ($\delta \gg g$), and as the system approaches resonance ($\delta\to0$), the dependence of interest becomes weaker. In Sec.~5, we turn to three-level atoms, for which the specific parameter values affect the system dynamics, and fix $\omega_A=1.2, \omega=1$ and $g=0.1$. However, additional calculations indicate that varying the parameters while maintaining the same ratio $\delta/g$ leaves the final results essentially unchanged.

\section{The Initial State without Photons}

In this section, we investigate the case in which the environment is initially in the vacuum state, and the system is prepared in a pure state. Our analysis concentrates on the Haar-random states and samples based on them. In Appendix A, we consider another special set of states discussed in the literature \cite{Entanglement, Sudden}.

The assumption of an initially photon-free environment implies that only the $\ket{e0}\leftrightarrow\ket{g1}$ oscillation occurs in each subsystem. The entropy is averaged over a large number of equally spaced time points within the corresponding oscillation period.

For convenience of comparison with the results of Ref.~ \cite{Does}, in this section, we use Meyer-Wallach measure (originally introduced in Ref~\cite{Wallach-Meyer}), represented here in the form
 \begin{equation}
		Q=N-\sum_{i=1}^{N}Tr\rho_i^2,
		\label{eq:ref50}
	\end{equation} 
  where $\rho_i$ is the reduced density matrix of the $i$-th atom. 

 \subsection{The system in Haar-random states}

In this subsection, we consider atomic systems initially prepared in Haar-random pure states generated using the Gurvitz parametrization described in Appendix B. For these ensembles, we compute and analyze the average dependence of $S_t$ on the initial entanglement quantified by $Q$. 

\begin{figure}[h]
\begin{minipage}[h]{0.48\linewidth}
\center{\includegraphics[width=1\linewidth]{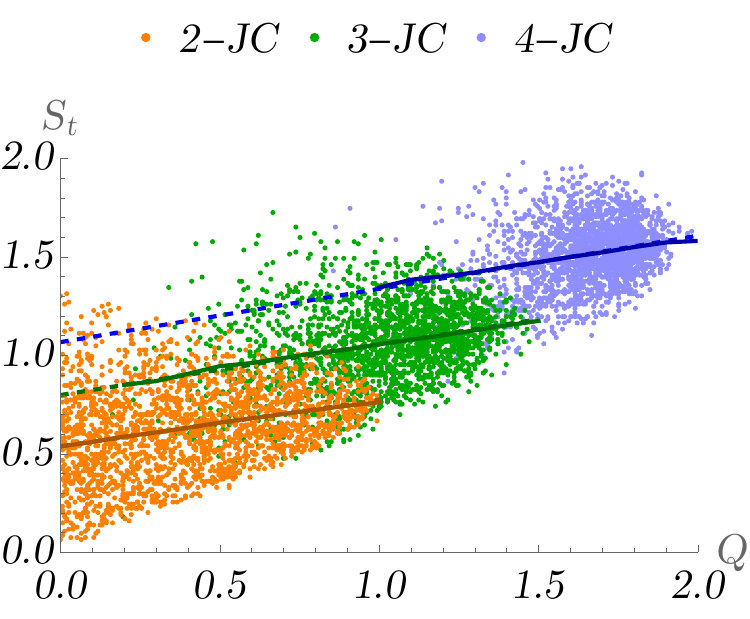}} (a) \\
\end{minipage}
\hfill
\begin{minipage}[h]{0.48\linewidth}
\center{\includegraphics[width=1\linewidth]{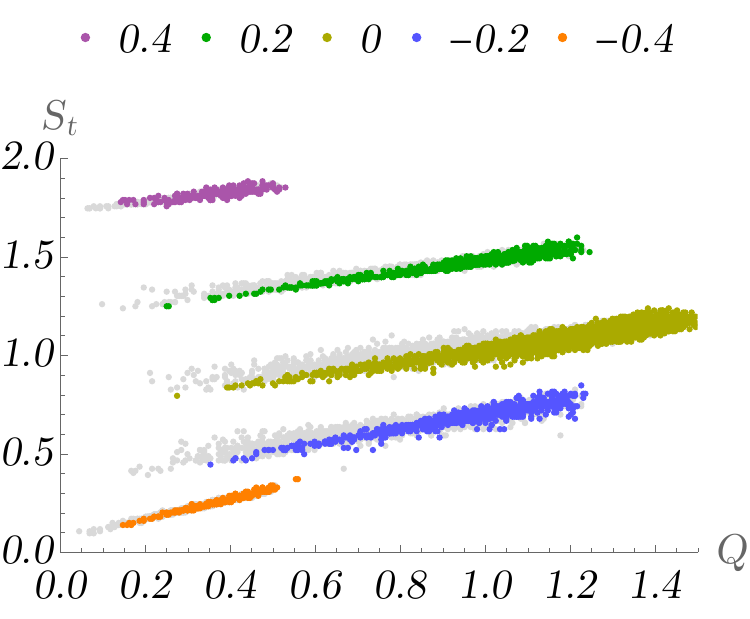}} (b)  \\
\end{minipage}
    \caption{Time-averaged entropy versus initial entanglement, quantified by the Meyer-Wallach measure. (a) The atomic systems are initially prepared in Haar-random pure states. Different colors correspond to different numbers of subsystems. The average curves (solid lines) closely follow the fitting straight lines (dashed). (b) Initial system states have different fixed values of  $\left\langle {E_i} \right\rangle$, indicated by different colors. Results for samples with the corresponding values of  $\left\langle {E} \right\rangle$ are shown in light gray. The clouds corresponding to equal $\left\langle {E_i} \right\rangle$ and $\left\langle {E} \right\rangle$ substantially overlap.}
    \label{fig:figure1}
\end{figure}

The results for $2\text{-JC}, 3\text{-JC}, 4\text{-JC}$ cases are shown in Fig.~1(a). Solid curves represent average dependences obtained by averaging $S_t$ over ensembles of states whose entanglement $Q$ lies within intervals of width  $\Delta Q=0.1$. Dashed lines denote fitting lines. All data in this subsection are calculated for $10^5$ random states. The average dependencies are approximately linear; however, due to the substantial scatter of data for $2\text{-JC}$ and the comparably rapid concentration of both $S_t$ and $Q$ values around their means as $N$ increases, the Pearson correlation coefficients are relatively small: 0.28, 0.28, 0.24 for the $2\text{-JC}, 3\text{-JC}, 4\text{-JC}$ cases, respectively. Here and below, all quantities are rounded to two decimal places, except for the slope angles.

The slope angles of the fitting lines are $13^\circ, 14^\circ, 14^\circ$ ($S_t$ and $Q$ are considered in the same scale). Throughout, slope angles are rounded to the nearest degree.
The obtained values are slightly smaller than those reported in Ref.~\cite{Does}, where the environment was infinite, and the analysis focused on the limit $t\rightarrow\infty$, yielding angles of $16^\circ,17^\circ$ and $18^\circ$.

The positivity of the average dependence of $S_t$ on $Q$ could be anticipated by analogy with the results of Ref.~\cite{Does}. Although the analytical justification for the $N\text{-JC}$ model is somewhat different, it likewise employs the concentration-of-measure phenomenon as an important ingredient.

In contrast to the model considered in Ref.~\cite{Does}, here it is not possible to choose a basis in which the phases of the expansion coefficients do not affect the entropy, such that the growth of its average value could be attributed solely to the equalization of the moduli of the coefficients as the system size increases. On the contrary, in the present analysis, the inclusion of phase differences plays a decisive role.

We work on the product basis constructed from the states $\ket{gm}$ and $\ket{em}$ of each subsystem, in which the diagonal elements of the density matrix are independent of relative phases. The emergence of nonzero phase differences leads to partial decoherence: different phase factors acquired by the terms contributing to the same off-diagonal element of the density matrix reduce its magnitude. A reduction in the magnitudes of the off-diagonal elements of an individual atom at $t=0$ contributes to an increase in the initial entanglement, whereas a reduction in the magnitudes of the off-diagonal elements of the system’s density matrix at later times contributes to an increase in the entropy. For any fixed set of coefficient moduli, this results in a simultaneous increase of both entanglement and entropy, effectively driving their values in a direction that yields a positive correlation between the two quantities. 

Moreover, the more homogeneous the moduli of the coefficients, the larger the effect of partial decoherence. This favors the positivity of the average dependence; however, it is still not sufficient to establish it without quantitative analysis.

On the other hand, according to the concentration-of-measure phenomenon, larger entanglement depths imply more homogeneous coefficient moduli distributions. Consequently, the quantity $\overline{S}_t$ for samples with higher entanglement depth is larger than for samples with lower depth (e.g., separable ones) even when systems include the same number of subsystems.  As a result, one observes a positive dependence of $\overline{S}_t$ on $Q$.

Although we employ the concentration-of-measure phenomenon in the analytical argument, it would be too strong to claim that it is itself the origin of the positivity of this dependence. Nevertheless, as in Ref.~\cite{Does}, the concentration-of-measure phenomenon can be said to govern the sensitivity of the fitting-line slope to the system size; however, the variation of the angle is very small.

To complement the qualitative analysis, we now consider the fractional contribution of the initial entanglement to the entropy growth 
 \begin{equation}
 \eta^{ent}=(\overline{S}_t-\overline{S}_t(Q=0))/\overline{S}_t,
 \label{eta}
  \end{equation} 
where $\overline{S}_t(Q=0)$ denotes the value of $S_t$ on the fitting line at $Q=0$.  As the system size increases, $ \eta^{ent}$ grows: for the $2\text{-JC}, 3\text{-JC}, 4\text{-JC}$ cases we obtain values $0.15, 0.23, 0.28$, respectively. These values exceed the corresponding values $0.08, 0.13, 0.16$ reported in Ref.~\cite{Does}. Using the ensemble-averaged entropy of maximally entangled states instead of Haar-random states in Eq.~(\ref{eta}) gives fractional contributions of $0.30, 0.32, 0.33$. These values emphasize the significant role of the initial entanglement. Notably, for the $2\text{-JC}, 3\text{-JC}, 4\text{-JC}$ cases, we find $\overline{S}_t(Q_{max})-\overline{S}_t(Q=0)=0.23, 0.36, 0.51$, respectively, whereas for initially separable systems adding one more subsystem increases the entropy by $0.27$. This indicates that the effect of the initial entanglement is comparable to or even larger than that of increasing the number of subsystems by one.

\subsection{Samples with energy constraints}

 Using the example of the states $\psi_1$ (considered in Appendix A), one can see that some samples show markedly different behavior from that in Fig.~1(a). The same was noticed in Ref.~\cite{Does} for states with a fixed average excited-state population. For Hamiltonian (\ref{Ham}), it corresponds to fixing the average energy of each atom. We denote the average energy of each atom, normalized to $\omega_A$, as $\left\langle {E_i} \right\rangle$. We consider $\left\langle {E_i} \right\rangle=-0.4, -0.2, 0, 0.2, 0.4$. 

Fig.~1(b) shows the results for the $3\text{-JC}$ case. The slope angles of the fitting lines, Pearson correlation coefficients, and fractional contributions of the initial entanglement for the  $2\text{-JC}$ and $3\text{-JC}$ cases are given in Appendix C, as well as the precision of $\left\langle {E_i} \right\rangle$ values.

For each sample with a given $\left\langle {E_i} \right\rangle$, a positive correlation is observed. A positive correlation for similar samples and $\sigma_x$-interactions was also reported in Ref.~\cite{Does} (whereas for $\sigma_z$ it was negative). This behavior can be attributed to the energy transfer between the system and the environment.
 However, unlike in Ref.~\cite{Does}, the results are not symmetric: for example, values of $S_t$ for states with $\left\langle {E_i} \right\rangle=0.4$ are larger than for states with $\left\langle {E_i} \right\rangle=-0.4$. This asymmetry is related to the role of the initial state of the environment: the closer the atomic states are to $\ket{g}$, the smaller their ability to become entangled with an initially photonless environment.

The contribution of the initial entanglement to the entropy is most significant for system states characterized by low values of $\left\langle {E_i} \right\rangle$. For example, for $3\text{-JC}$ at $\left\langle {E_i} \right\rangle=-0.4$ the fractional contribution $\eta^{ent}$ is $0.76$, while for $\left\langle {E_i} \right\rangle=0.4$ it is $0.05$.

A modest contribution to the change in $\eta^{ent}$ comes from the increase of the fitting-line slope at low $\left\langle {E_i} \right\rangle$. Notably, for both the $2\text{-JC}$ and $3\text{-JC}$ cases at $\left\langle {E_i} \right\rangle=-0.4$ the angle reaches $28^\circ$, which is significantly larger than in the Haar-random case. 
It happens because the entropy of states with smaller $\left\langle {E_i} \right\rangle$ is more sensitive to phase differences. To understand the origin of this effect, it is convenient to consider the $2\text{-JC}$ case and examine the density matrix of the two-atom system. 

We consider the atomic system in the basis $\ket{ee},\ket{eg},\ket{ge},\ket{gg}$ and use the parametrization (\ref{H1}). In what follows, by phase factors, we will specifically mean those associated with 
$\varphi_i$, rather than those arising from time evolution. Since at any time each subsystem contains at most one excitation, tracing out the field degrees of freedom gives rise to terms with two distinct phase factors only in the density-matrix elements $\ket{gg}\bra{ge}, \ket{gg}\bra{eg},\ket{ge}\bra{gg}, \ket{eg}\bra{gg}$. In other elements, all terms have identical phase factors, so that the moduli of these elements are unaffected by the inclusion of phase differences.
 
 If the initial $\langle E_i \rangle$ is low, the density-matrix elements $\ket{ee}\bra{ij}$ and $ \ket{ij}\bra{ee}$ are suppressed, and their contribution to the entropy becomes negligible. In this case, the remaining matrix elements play a more significant role. Consequently, variations associated with the inclusion of phase differences have a stronger impact on the entropy than in the case where all density-matrix elements have comparable magnitudes, which occurs for initial states with $\langle E_i \rangle$ close to unity.  At the same time, the effect of introducing phase differences on the initial entanglement is the same for two states that are related by the exchange $\ket{e}\leftrightarrow \ket{g}$. Therefore, the stronger sensitivity of entropy to phase differences for states with small $\langle E_i \rangle$ implies a larger slope angle. For low-energy initial states, intrasystem entanglement turns out to be the primary driver of entropy growth.

Similar trends are observed for samples of states with fixed values of the total mean energy of the system (i.e., summed over all qubits and divided by the number of qubits) normalized to $\omega_A$. As illustrated in Fig.~1(b), the point clouds corresponding to equal $\langle E_i \rangle$ and $\langle E \rangle$ exhibit significant overlap, but in the latter case, the scatter in $S_t$ for low entanglement is slightly larger. Slope angles, Pearson coefficients, and fractional contributions $\eta^{ent}$ are listed in Appendix C.

What is particularly notable in Fig.~1(b) and the tables in Appendix C is that, for both types of samples considered in this subsubsection, the dependencies are more pronounced than for Haar-random states, as indicated by the Pearson coefficient approaching unity. A simple argument suggesting this trend can be given by considering the separable limit. 
For a single atom, the entropy depends only on the squared modulus of the coefficient of $\ket{e0}$ in the expansion of the system state, which also determines the value of $\langle E_i \rangle$. 
Therefore, in separable systems, all states with fixed $\left\langle {E_i} \right\rangle$ yield well-defined values of $S_t$. Including states with different values of $\left\langle {E_i} \right\rangle$ across atoms results in a scatter in $S_t$. Nevertheless, fixing $\left\langle {E} \right\rangle$ significantly constrains the spread.

So far, we have considered only two-level atoms, with the environment initially prepared in the vacuum state. In the following sections, we consider environments with initial photons and three-level atoms. We will not return to samples with fixed $\left\langle {E_i} \right\rangle$ or $\left\langle {E} \right\rangle$. However, here we emphasize that in all such cases the average dependence of $S_t$ on the initial entanglement remains positive and approximately linear, with Pearson coefficients much closer to unity than for Haar-random ensembles. Thus, for the samples considered in this subsubsection, the dependence of interest is much closer to being truly linear.

\section{Effect of the Initial Photon Number}
	
In the previous section, we concentrated on cases in which the initial field states were $\ket{0}$. We now assume that each subsystem initially contains $n$ photons, where $n$ is a fixed integer. In this scenario, two types of transitions occur within each subsystem: $\ket{gn}\leftrightarrow\ket{e(n-1)}$ and $\ket{en}\leftrightarrow\ket{g(n+1)}$. Oscillations, associated with the above transitions, generally have distinct, incommensurate periods. Therefore, in our calculations, we average over many time points within sufficiently long time intervals. 

In this section, we restrict our analysis to systems composed of two subsystems. To facilitate comparison with the results presented in the following section, we use the entanglement entropy, defined as the von Neumann entropy in base two of the reduced density matrix of one atom, as a measure of entanglement. 

\begin{figure}
\begin{minipage}[h]{0.48\linewidth}
\center{\includegraphics[width=1\linewidth]{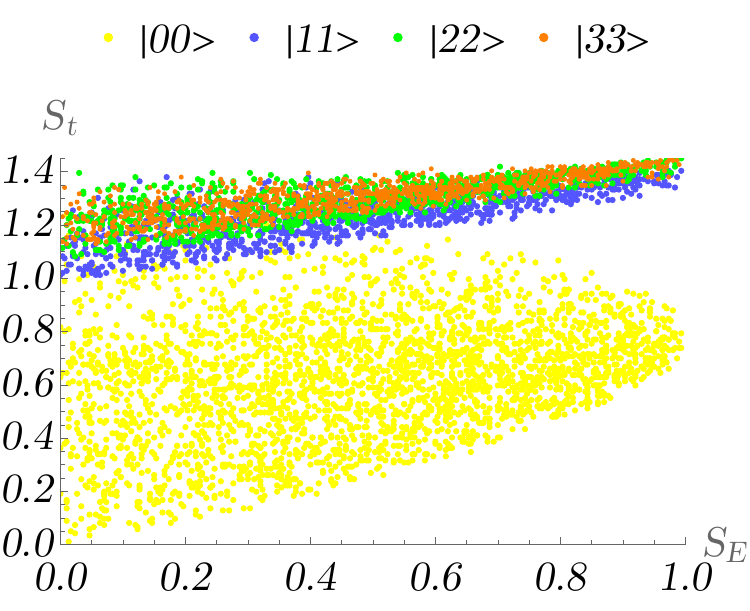}} (a) \\
\end{minipage}
\hfill
\begin{minipage}[h]{0.48\linewidth}
\center{\includegraphics[width=1\linewidth]{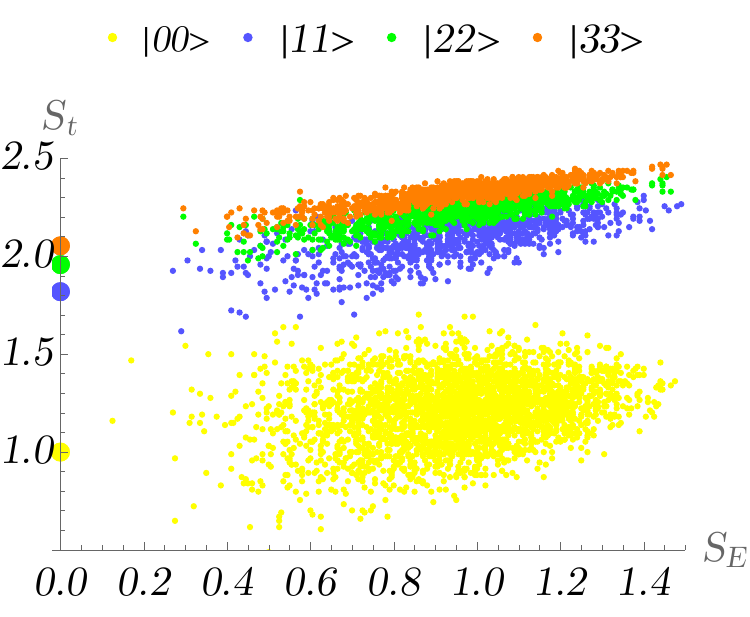}} (b)  \\
\end{minipage}
    \caption{Time-averaged entropy versus initial entropy of entanglement for pure Haar-random initial system states and different initial photon numbers (indicated by different colors). The system consists of atoms with (a) a ground and a single excited level, and (b) a ground and two excited levels. The bold markers on the $S_t$ axis indicate the mean values for ensembles of random separable states.} 
    \label{fig:figure3}
\end{figure}

The participation of the $\ket{g}$ component of the initial atomic state in the oscillatory dynamics leads to larger time-averaged entropy values and a stronger concentration of data points around the fitting lines. The slight increase in $S_t$ with increasing $n$ can also be attributed to the enhancement of the effective system–environment coupling strength (described by the off-diagonal elements of the Hamiltonian (\ref{Ham}) that are proportional to the square root of the number of excitations in the subsystem). These trends are illustrated in Fig.~2(a).

Here and in the following section, the Pearson correlation coefficients, slopes, and fractional contributions are obtained by averaging over ensembles of $10^4$ randomly generated states for $n=1, 2, 3$ and of $5\cdot10^4$ for $n=0$. For $n=0, 1, 2, 3$, the Pearson correlation coefficient takes the values $0.28, 0.79, 0.81, 0.84$, respectively. Therefore, for $n>0$, discussing the dependence of the time-averaged entropy on the initial entanglement becomes more meaningful. However, $\eta^{ent}$ decreases, taking the values $ 0.18, 0.10, 0.09, 0.08$, respectively.  This behavior is not solely due to the increase of $\overline{S}_t$. The slope angle of the fitting line takes values $14 ^{\circ}, 15^{\circ}, 14^{\circ}, 13^{\circ}$, decreasing for $n>1$. We return to a discussion of this trend in Sec. 6, but here we note that it resembles the situation where, during the interaction of a small system with a large environment, the memory of its initial state is progressively lost—in our specific case, the memory of the internal entanglement.
At the same time, in the model considered in Ref. \cite{Does}, the environment was assumed to be infinite, yet the slope of the fitting line in all cases remained noticeably different from zero.

\section{Systems with Three-Level Atoms}

We now consider generalisations of the Jaynes-Cummings model to systems with three-level atoms. Such generalisations have been discussed previously \cite{cascade,Inf}, although less extensively. Ref. \cite{cascade} investigated the population dynamics of a three-level atom interacting with a single field mode, depending on the atom’s initial state. In that work, the cascade configuration was considered, where transitions can occur either between the first and second excited levels or between the first excited and ground levels. In this subsection, instead of the Hamiltonian (\ref{Ham}) for each subsystem, we adopt the Hamiltonian of Ref.~\cite{cascade}:

\begin{equation}
\begin{matrix}
H=
-\omega_A\ket{g}\bra{g}+ 0 \ket{e1}\bra{e1}+\omega_A\ket{e2}\bra{e2}
+\omega(a^\dagger a+1/2)\\
+g\big( a(\ket{e1}\bra{g}+\ket{e2}\bra{e1})+a^\dagger(\ket{g}\bra{e1}+\ket{e1}\bra{e2})\big),
\end{matrix}
\label{Ham2}
\end{equation} 
where $\ket{e_1}$ and $\ket{e_2}$ are the first and second excited levels of the atom, and $\omega_A$ is the energy gap between $\ket{e_2}$ and $\ket{e_1}$ levels, as well as between the $\ket{e_1}$ and $\ket{g}$ levels.

Since the Meyer–Wallach measure is defined only for qubits \cite{Wallach-Meyer}  and not for qutrits, here we employ an alternative entanglement measure. For simplicity, we restrict our consideration to two subsystems, which allows us to quantify entanglement via the entanglement entropy.

For $n=0, 1, 2, 3$ results are shown in Fig. 2(b); the slope angles of the fitting lines are $12^\circ, 17^\circ, 16^\circ, 15^\circ$. The corresponding Pearson correlation coefficients are $0.26, 0.63, 0.83, 0.89$. As in the case of two-level atoms, the slope of the fitting line is maximal for $n = 1$, and the Pearson correlation coefficient increases with $n$.

The contribution of the initial entanglement $\eta_{ent}$ takes the values $0.17, 0.13$, $0.12, 0.11$. A comparison with the results of the previous section shows that switching to qutrits slightly increases $\eta_{ent}$ for $n>0$, whereas for $n=0$ it slightly decreases.

 \section{Mixed States}
So far, we have considered only pure states; however, most physical states are mixed. In this section, we focus on ensembles of Haar-random mixed states, which we generate using the Ginibre matrix method (see Appendix D). We restrict ourselves to the simplest case of a two-qubit system, which allows the use of concurrence as an entanglement measure. Concurrence is defined as   
\begin{equation}
C=\max\left\{0,\sqrt{\lambda_1}-\sqrt{\lambda_2}-\sqrt{\lambda_3}-\sqrt{\lambda_4}\right\},
\end{equation}
where $\lambda_i$ are the eigenvalues of $\rho (\sigma_y \otimes\sigma_y )\rho^* (\sigma_y \otimes\sigma_y )$, $\rho$ is the density matrix of the system, and $\sigma_y$ is the corresponding Pauli operator (see, e.g., Ref~\cite{Wootters}).

The average dependence of $S_t$ on the concurrence is negative: for $n=0, 1, 2$ the slope angles of the fitting line take the values $-25^\circ, -11^\circ, -8^\circ$, correspondingly. These values were obtained from ensembles of $5\cdot10^4$ random mixed states.
 However, mixed states always possess a nonzero initial entropy  $S_{in}$. The higher the initial mixedness, the lower the intrinsic entanglement of the system (the inverse relationship between entropy and entanglement in mixed two-qubit states was discussed in Refs.~\cite{versus, Quantum}). Thus, when considering the change in entropy, $\Delta S_t=S_t-S_{in}$, a positive correlation is again observed, as shown in Fig.~3(a).

\begin{figure}
\begin{minipage}[h]{0.48\linewidth}
\center{\includegraphics[width=1\linewidth]{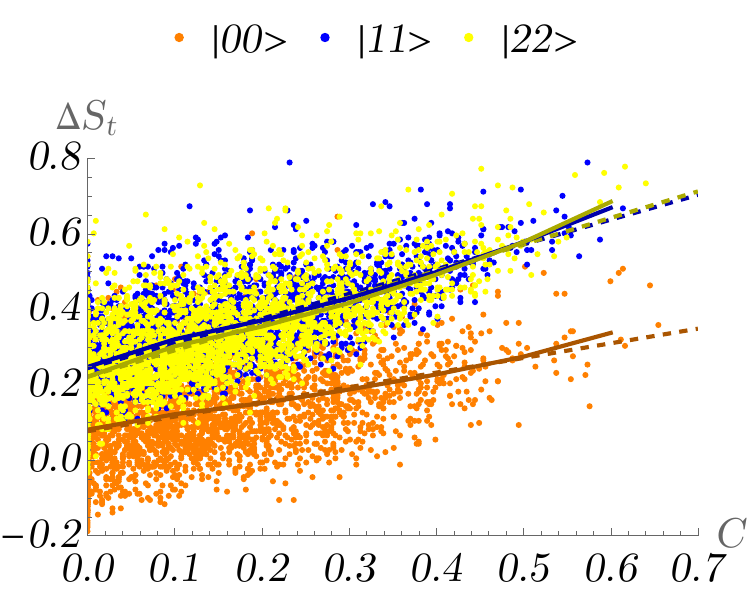}} (a) \\
\end{minipage}
\hfill
\begin{minipage}[h]{0.48\linewidth}
\center{\includegraphics[width=1\linewidth]{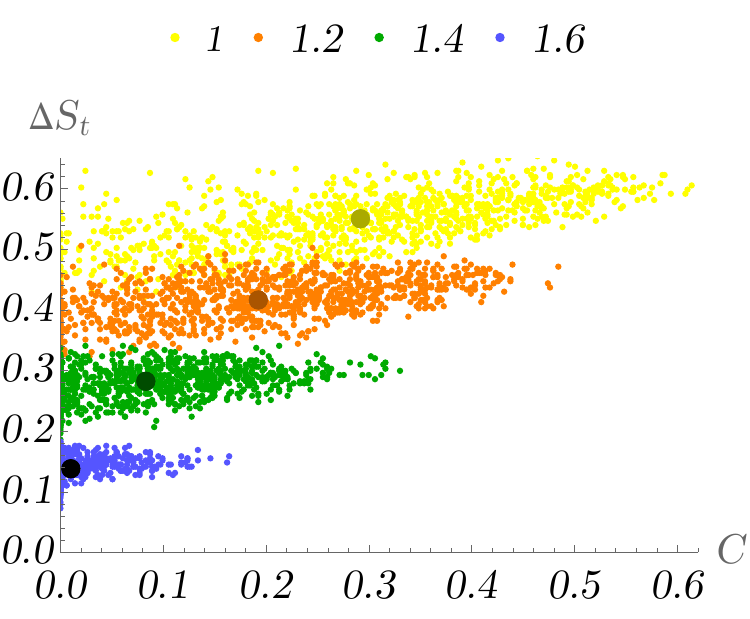}} (b)  \\
\end{minipage}
    \caption{Time-averaged entropy change versus initial concurrence. (a) For initial Haar-random mixed states of the system and for different environment states (indicated by different colors). Solid lines correspond to the average curves, while dashed lines represent the fitting lines. (b) For samples with fixed initial mixedness of the system, measured by von Neumann entropy. Different colors correspond to different mixedness values. The bold dots indicate ensemble averages. The environment is initially in the state $\ket{22}$.} 
    \label{fig:figure3}
\end{figure}
  
 In Fig.~3(a), we see that for many states with small entanglement values, $\Delta S_t$ is negative, indicating that the entropy decreases rather than increases. For example, if the initial system state is maximally mixed, the interaction with the environment can only reduce the system’s entropy. At the same time, for non–maximally mixed initial states, the cases $n>0$ provide much greater potential for system–environment entanglement; therefore, negative $\Delta S_t$ values are mainly observed for  $n=0$.

The cases $n = 1$ and $n = 2$ involve the same number of oscillatory processes but differ in the effective coupling strength. In contrast to Sec. 4, here larger values of $n$ do not lead to larger values of $S_t$. This apparent discrepancy can be explained by the role of the initial mixedness of the individual atomic states. For nearly pure initial states of individual atoms, an increase in the effective coupling strength enhances atom–environment entanglement. For more mixed states, stronger coupling promotes not only higher mixing, but also a stronger purification during the evolution. As a result, beyond a certain degree of mixedness, increasing the interaction strength on average suppresses the growth of entropy. 

In ensembles of pure states, lower entanglement corresponds to higher purity of individual atomic states. Therefore, for less entangled states, an increase in $n$ is accompanied by a larger increase in $S_t$, leading to a reduction in the slope of the fitting lines in  Fig.~2. For ensembles of mixed states, an inverse correlation between the concurrence and the mixedness of individual atoms leads to a slightly larger slope for $n = 2$ than for $n = 1$, as observed in Fig.~3(a). 

We characterize the dependencies using the slopes of the fitting lines, although those for mixed states deviate significantly from linear behavior, in contrast to the dependencies for pure states. This effect is particularly pronounced for large $C$: maximal entanglement can only be achieved by pure states, the values of $\overline{S}_t(C_{max})$ must coincide for mixed and pure ensembles, forcing a sharper increase in entropy. However, the fraction of strongly entangled states is relatively small.

  The slope angles for random mixed states are $21^\circ, 33^\circ, 35^\circ$ for $n=0, 1, 2$, respectively. Fractional contributions are $0.39, 0.25, 0.29$. These values were obtained from ensembles of $5\cdot10^4$ random mixed states. Both quantities are significantly larger than those for random pure states. However, this effect primarily results from variations in the degree of mixedness of the system, since larger initial mixedness implies not only lower entanglement, but also a smaller potential for entropy growth. Considering system states with fixed initial entropy values, we observe that both concurrence and time-averaged entropy decrease with increasing initial mixedness. Specifically, for $n=2$, this trend can be seen in Fig.~3(b). For technical details, see Appendix E.  
 
 For ensembles with fixed initial mixedness, the slopes of the fitting lines are smaller than for random mixed or pure states: $10^{\circ}, 8^{\circ}, 6^{\circ}, 6^{\circ}$ for $S_{in}=1, 1.2, 1.4, 1.6$, respectively. Therefore, as the initial mixedness decreases, the slope increases, taking the value of $15^{\circ}$ for random pure states (in Section 4, for entanglement entropy, we obtained $14^\circ$).

\section{Conclusions}

We examined how the initial internal entanglement of atomic systems affects the time-averaged entropy generated via their interaction with a photonic environment in several generalizations of the Jaynes–Cummings model. In agreement with previous findings, we found that for samples of pure Haar-random initial states, their entanglement, on average, contributes to the time-averaged entropy. Although this contribution is modest for Haar-random states, for low-energy ensembles it can even become dominant.
 
 In this work, we extended previous results by investigating the influence of the initial state of the environment. We found that, for systems in pure Haar-random states, increasing the initial photon number (for $n>1$), on average, decreases the fractional contribution of the initial entanglement to the entropy growth. In addition to pure states, we considered Haar-random mixed samples and noticed that although their fitting slopes are larger, restricting the samples to a fixed degree of mixedness yields slopes smaller than those of the pure-state ensembles.

These results help clarify the role of intrasystem correlations in quantum-informational processes. At the same time, some questions remain open; for example, how the inclusion of interactions between subsystems or many-body interactions affects the dependence, and whether signatures of initial internal correlations persist in models with continuous degrees of freedom. Moreover, the present results provide a useful starting point for future studies of whether initial intrasystem entanglement influences information copying and proliferation in quantum Darwinism scenarios.

\section*{Acknowledgement}
The author acknowledges the use of ChatGPT (OpenAI) for assistance with translation and stylistic editing. All scientific content is the author's own.

\appendix

\section*{Appendix A. States with One Atomic Excitation}
\renewcommand{\theequation}{A.\arabic{equation}}
Here we consider states of the type:
 \begin{equation}
  \begin{matrix}
\ket{ \psi_1(0)}=\cos \alpha \ket{e0}\ket{g0}+\sin \alpha \ket{g0}\ket{e0}.
 \end{matrix}
 \label{alpha}
   \end{equation}
 For the double Jaynes-Cummings model, the dynamics of such states were analyzed in Refs. \cite{Entanglement}. The degree of entanglement is determined by the value of $\alpha$. In the basis $\ket{ee},\ket{eg},\ket{ge},\ket{gg}$, the density matrix of the atomic system at any time has the form
     \begin{equation}
\begin{pmatrix}
0&0&0&0\\
0&x_1(t) \cos^2\alpha&x_1(t) \cos\alpha \sin\alpha&0\\
0&x_1(t) \cos\alpha \sin\alpha&x_1(t) \sin^2\alpha&0\\
0&0&0&x_2(t)
		\end{pmatrix},
		\label{eq:2}
		\end{equation}
where $x_1(t),x_2(t)$ are expressed through $g$ and $\delta$ and do not depend on $\alpha$. Varying $\alpha$ corresponds to a transition between bases that does not change entropy. Introducing a phase factor, $\psi'_1(0)=\cos\alpha \ket{e0}\ket{g0}+e^{i \phi}\sin\alpha \ket{g0}\ket{e0}$ also leaves the entropy unchanged; $\psi'_1$ includes all states of $2\text{-JC}$ that have exactly one atomic excitation, which provides a clear physical interpretation of this ensemble.

This result can be generalized to systems consisting of  $N$ subsystems. For initial states of the form (with complex $m_i$):
 \begin{multline}
 \ket{\psi_1^{(N)}(0)}=m_{1} \ket{e0}\ket{g0}\ldots \ket{g0}+m_{2} \ket{g0}\ket{e0}\ket{g0}\ldots \ket{g0} +\dots+ m_{N} \ket{g0}\ldots\ket{g0}\ket{e0}
 \end{multline}
 at any moment of time, the entropy does not depend on $m_i$ and, consequently, on the initial entanglement.

\section*{Appendix B. Generation of Pure Haar-Random States}
\renewcommand{\theequation}{B.\arabic{equation}}
\setcounter{equation}{0}
\label{random}
Haar-random pure states are obtained by applying an arbitrary unitary transformation $U$, whose explicit form is given in the Appendix of Ref.~\cite{Composed}, to a fixed reference vector $(1, 0, …, 0)^T$. In our calculations, we use the corresponding parametrized state vector in an $l = 2^N$-dimensional Hilbert space:
\begin{equation}
\ket{\psi}=
\begin{pmatrix}
			\cos \vartheta_{l-1}\\
			\sin \vartheta_{l-1} \cos \vartheta_{l-2} e^{i \varphi_{l-1}}\\
			\dots\\
			(\prod_{k=2}^{l-1} \sin \vartheta_k) \cos \vartheta_1 e^{i \varphi_2}\\
			(\prod_{k=1}^{l-1} \sin \vartheta_k) e^{i \varphi_1}	
			\end{pmatrix}.
			\label{H1}
	\end{equation} 	
Here angles $\vartheta_k\in[0,\pi/2]$ are taken with the probability density $P(\vartheta_k)=k \sin(2\vartheta_k)(\sin \vartheta_k)^{2k-2}$ and  $\varphi_k$ are uniformly distributed over the interval $[0, 2\pi)$.

\section*{Appendix C. Results for Samples with Energy Constraints and the Precision of $\langle E_i \rangle$ and $\langle E \rangle$ Values}
  
   \begin{table}[h]
 \centering
 \caption{Slope angles, Pearson correlation coefficients, and fractional contributions of entanglement for samples of system states with fixed $\langle E_i \rangle$.}
 \begin{tabular}{c c c c c c}
 $\langle E_i \rangle$ & $-0.4$ & $-0.2$ & $0$ & $0.2$ & $0.4$ \\
 Angle ($2\text{-JC}$)   & $28^\circ$ & $21^\circ$ & $16^\circ$ & $16^\circ$ & $14^\circ$ \\
 Angle ($3\text{-JC}$)   & $28^\circ$ & $21^\circ$ & $18^\circ$ & $17^\circ$ & $14^\circ$ \\
 Pearson ($2\text{-JC}$) & 1.00 & 0.99 & 0.94 & 1.00 & 1.00 \\
 Pearson ($3\text{-JC}$) & 0.98 & 0.94 & 0.93 & 0.97 & 0.80 \\
  $\eta^{ent}$ ($2\text{-JC}$)&  0.72 & 0.47 & 0.28 & 0.15 & 0.04\\
  $\eta^{ent}$ ($3\text{-JC}$)&  0.76 & 0.53 & 0.36 & 0.19 & 0.05\\
 \end{tabular}
 \end{table}
  
To obtain the results shown in Table 1, states were selected from the set of Haar-random states by rounding $\langle E_i \rangle$ to 0.005 for the $2\text{-JC}$ case. 
For the $3\text{-JC}$ case, $\langle E_i \rangle$ was rounded to 0.01 at $\langle E_i \rangle = 0, \pm 0.2$ and to 0.02 at $\langle E_i \rangle = \pm 0.4$.

 \begin{table}[h]
\centering
\caption{Slope angles, Pearson correlation coefficients, and fractional contributions of entanglement for samples of system states with fixed $\langle E \rangle$.}
\begin{tabular}{c c c c c c}
$\langle E \rangle$ & $-0.4$ & $-0.2$ & $0$ & $0.2$ & $0.4$ \\
Angle ($2\text{-JC}$)   & $28^\circ$ & $19^\circ$ & $12^\circ$ & $15^\circ$ & $14^\circ$ \\
Angle ($3\text{-JC}$)   & $26^\circ$ & $18^\circ$ & $14^\circ$ & $15^\circ$ & $15^\circ$ \\
Pearson ($2\text{-JC}$) & 1.00 & 0.97 & 0.87 & 1.00 & 1.00 \\
Pearson ($3\text{-JC}$) & 0.99 & 0.94 & 0.92 & 0.91 & 0.98 \\
  $\eta^{ent}$ ($2\text{-JC}$)&  0.62 & 0.33 & 0.16 & 0.10 & 0.03\\
  $\eta^{ent}$ ($3\text{-JC}$)&  0.67 & 0.40 & 0.25 & 0.15 & 0.05\\
\end{tabular}
\end{table}

To obtain the results shown in Table 2, states were selected from the set of Haar-random states for the $2\text{-JC}$ case by rounding $\langle E \rangle$ to $10^{-4}$ at $\langle E \rangle = 0, \pm 0.2$ and to $2 \cdot10^{-4}$ at $\langle E \rangle = \pm 0.4$. 
For the $3\text{-JC}$ case, $\langle E \rangle$ was rounded to $10^{-5}$ at $\langle E \rangle = 0$, to $5\cdot10^{-5}$ at $\langle E \rangle =\pm 0.2$ and to 0.01 at $\langle E\rangle = \pm 0.4$. For both types of samples, ensemble sizes ranged from several hundred to several thousand states, and the results were insensitive to further increases in the ensemble size.

\section*{Appendix D. Ginibre Matrix Method}
\renewcommand{\theequation}{D.\arabic{equation}}
Random mixed states were generated using Ginibre matrices M, i.e., complex matrices whose real and imaginary components are sampled independently from a normal distribution with zero mean and unit variance. Normalized positive matrices were then constructed as 
\begin{equation}
\rho=\frac{M M^\dagger}{Tr(M M^\dagger)}.
\end{equation}
 This method ensures a Haar-random distribution over the space of mixed states \cite{Generating}.

\section*{Appendix E. The precision of initial mixedness values for ensembles with fixed $S_{in}$}
To obtain Fig.~3(b) and related data, we selected from the Haar-random mixed states those with von Neumann entropy values of 1 and 1.6 with an accuracy of 0.002, and those with entropy values of 1.2 and 1.4 with an accuracy of 0.0005. The choice of $S_{in}$ values and their precision was motivated by the need to obtain sufficient statistics. The slope angles were calculated from the ensembles containing on the order of $10^4$ states, and the results were insensitive to further increases in the ensemble size.

\end{document}